\documentclass[preprint,12pt]{elsarticle}
\usepackage{amsfonts,amssymb,amsmath}
\usepackage{slashed}
\journal{Physics Letters A}

\begin{document}

\begin{frontmatter}

\title{The Rarita-Schwinger Particles Under the Influence of Strong Magnetic Fields}
\author[UFSC]{M.G. de Paoli}\ead{marcelodepaoli@gmail.com}
\author[UFSC]{L.B. Castro}\ead{luis.castro@pgfsc.ufsc.br}
\author[UFSC]{D.P. Menezes}\ead{debora.p.m@ufsc.br}
\author[UFSC]{C.C. Barros Jr.}\ead{barros@fsc.ufsc.br}

\address[UFSC]{Departamento de F\'{\i}sica - CFM - Universidade Federal de Santa
Catarina, Florian\'opolis - SC - CP. 476 - CEP 88.040 - 900 - Brazil.}

\begin{abstract}
In this work, we calculate the solutions of the
Rarita-Schwinger equation with the inclusion of the
electromagnetic interaction. Our gauge and coupling
prescription choices lead to Dirac-type solutions.
One of the consequences of our results is that all
inconsistencies related to noncovariance and noncausality are avoided
and the Landau
level occupation of particles turns up to be quite different from the
usual spin 1/2 particle system occupation numbers.
\end{abstract}


\end{frontmatter}

\section{Introduction}
It is well known that in heavy ion collisions, when dense matter is
formed, it is rich in delta resonances \cite{old_papers}.
Even if low energy physics is considered, this kind of resonance is
very important. In the $\pi N$ interactions for example, a
problem that has been studied for a long time, the $\Delta$ resonance plays a central
role \cite{eto,oo,manecoelho,mane}. The scattering amplitude, saturated
by the $\Delta$, is dominated by this contribution in the $\pi^+p$
channel and it is also important in the other ones. The importance of
spin 3/2 baryons is also observed for other hadron interactions, as in
hyperon interactions, where resonances with spin 3/2 ($\Sigma^*$(1385),
$\Xi^*$(1533) and others) \cite{cy} also dominate the scattering
amplitudes in many processes. Studies involving the chiral scalar form
factors of the nucleons \cite{mane} and hyperons also show the dominance
of the $\Delta$ and hyperon resonances.

Hadronic interactions are also important in high energy physics.
When considering particles produced in a close region, they may be
considered as comovers in an expanding system, and their relative
energy is small. Therefore, the effect of the resonances may have a
relevant contribution in the whole process.

The above mentioned facts were a strong motivation for the delta baryons to be considered as
 possible constituents of neutron stars \cite{glendenning, sergio,hua,
   veronica}. The appearance of particles in neutron stars
 described by relativistic models depends on the symmetry energy
 density dependence \cite{symmetry}, the choice of couplings between
baryons and interacting mesons (still unknown)
\cite{glendenning,sergio} and the possible appearance of pion and kaon
condensations \cite{glendenning, livro_glen, panda}.
In \cite{sergio} the contribution of the four delta particles
($\Delta^-,\Delta^0,\Delta^+$,$\Delta^{++}$) is taken into account in a
mean field theory (MFT) approach and the related
Rarita-Schwinger (RS) equation of motion \cite{rs}
is solved self-consistently with the other equations of motion that describe the
other baryons and mesons. The population of $\Delta$ particles in neutron stars for different asymmetries
($\beta$-equilibrium is not enforced) is then obtained. In \cite{hua}, the influence of $\Delta$ resonances on the
maximum star mass is investigated with two different relativistic models.
Indeed, the inclusion of $\Delta$ baryons leads to an increase of the maximum mass within the non-linear Walecka model \cite{nlwm}
and a decrease within a chiral hadronic model \cite{fst}.
Neutron stars with $\Delta$s are revisited in \cite{veronica} and they seem to suppress the hyperon abundances considerably.
In the last years, neutron stars subject to strong magnetic fields,
named magnetars by Duncan \cite{duncan},
have been a source of intense investigation \cite{magnetars}. If
one assumes that $\Delta$ particles have to be seriously treated as
possible constituents of these compact objects, the RS equation with
the inclusion of the magnetic field \cite{johnson, fradkin,mathews}
has to be taken into account.

Recently, the influence of magnetic fields on the QCD phase diagram for all possible temperatures and
chemical potentials has attracted some attention \cite{rapid,andersen}.
In \cite{rapid} one can see that, for fields greater than $10^{17} \,{\rm G}$
the critical end point and the first order coexistence region are affected.
One observes an increase on the size of the first order segment of the  transition line with $T_{\rm CEP}$ moving to
higher values as the magnetic field increases. This result is certainly of interest for the
physics of heavy ion collisions at intermediate energies, where huge fields, of the order $|eB| \ge m_\pi^2 \sim 10^{18} \, {\rm G}$,  can be created by heavy ion currents due to the spectator nucleons. The field intensity depends on the centrality and beam momentum so that  $|eB| \approx 5 m_\pi^2$ can be reached at RHIC while $|eB| \approx 15 m_\pi^2$ can be reached at the LHC.

In \cite{rhic} matter is described by free gases of baryons and mesons under the influence of a constant magnetic field and particle rations
and yields are calculated in such a way that the results for the
temperature and chemical potential are chosen for the minimum $\chi^2$ fit.
The results show that, even for the free Fermi and Boson gas models, a strong magnetic field plays an important role.
The inclusion of the magnetic field improves the data fit up to a field of the order of $B=10^{19}$ G. For stronger magnetic fields,
it becomes worse again. In this calculation, baryons with
spin 3/2 subject to an external magnetic field had to be described and
the degeneracy of the Landau levels (LL) was an important quantity.

The Rarita-Schwinger (RS) equation \cite{rs}
describes spin 3/2 particles and has been studied widely because of
 its intrinsic interest and also for applications in different
  research fields, and not only the above mentioned ones. The onus of
  inconsistencies represent a disadvantage to the RS equation.
The quantization of the interacting spin 3/2 field turned out to be
inconsistent with Lorentz covariance \cite{johnson}.
Moreover, the wave fronts of the classical solutions of the RS equation
present noncausal modes of propagation \cite{velo,hortacsu}. This kind of inconsistencies are
old problems and several remedies have been suggested through the years
\cite{pasca}-\cite{kru},  including an interesting prescription of a nonminimal
  coupling \cite{porrati}. From this last reference, it is clear that,
  at least for constant fields, the ones responsible for the
  noncausality problem \cite{velo,hortacsu} and the inconsistencies
  with Lorenz covariance \cite{johnson}, a valid, causal Langrangian
  exists.

For the reasons mentioned above, a complete solution for the RS equation under the influence
of magnetic fields deserves a detailed revision, which we perform in
the present work. We next follow the  approach developed in
  \cite{melrose} and \cite{greiner} and circumvent the  problems just
  mentioned, without having to rely on nonminimal couplings.
 To the best of our knowledge, this is the first
  time that a complete calculation of the Rarita-Schwinger equation
  with the inclusion of the magnetic interaction, including the
  occupation of the LL, is obtained in a simple way, avoiding the
  problems observed in older calculations.

\section{Formalism}

A spin-$3/2$ field can be described by the Rarita-Schwinger(RS) Lagrangian given by (with units in which $\hbar=c=1$)
\begin{equation}\label{lrsa}
    \mathcal{L}^{RS}(A)=\bar{\psi}^{\mu}\left( i\partial_{\alpha}\Gamma_{\mu}\,^{\alpha}\,_{\nu}(A)-mB_{\mu\nu}(A) \right)\psi^{\nu},
\end{equation}
\noindent where
$\bar{\psi}^{\mu}=\left(\psi^{\mu}\right)^{\dagger}\gamma^{0}$ is the
adjoint vector spinor, $m$ is the particle mass, $\psi^{\nu}$ is the
vector spinor and the matrices $\Gamma_{\mu}\,^{\alpha}\,_{\nu}$ and
$B_{\mu\nu}$ are given by \cite{nath3}:
\begin{equation}\label{ga}
    \Gamma_{\mu}\,^{\alpha}\,_{\nu}(A)=g_{\mu\nu}\gamma^{\alpha}+A\left(
      \gamma_{\mu} g_{\nu}^{\alpha}+ g_{\mu}^{\alpha} \gamma_{\nu} \right)+
    B\gamma_{\mu}\gamma^{\alpha}\gamma_{\nu}\,,
\end{equation}
\begin{equation}\label{ba}
    B_{\mu\nu}(A)=g_{\mu\nu}-C\gamma_{\mu}\gamma_{\nu}\,,
\end{equation}
\noindent with
\begin{equation}\label{b}
    B\equiv\frac{3}{2}\,A^{2}+A+\frac{1}{2}
\end{equation}
\begin{equation}\label{c}
    C\equiv3A^{2}+3A+1\,.
\end{equation}
\noindent where the metric tensor is $g_{\mu\nu}=\mathrm{diag}\left( 1,-1,-1,-1 \right)$, $\gamma^{\mu}$ are Dirac matrices and $A$ is an arbitrary parameter (except that $A\neq-1/2$).
The parameter $A$ has no physical meaning \cite{johnson,fradkin} and the conventional choices made in the literature are $A=-1/3$ that corresponds to the Lagrangian originally proposed in \cite{rs},
or, $A=0$ \cite{mathews}, or, for $A=-1$ the Lagrangian reduces to \cite{weinberg}
\begin{equation}\label{lrs}
 \mathcal{L}^{RS}(A=-1)=-\frac{i}{2}\,\bar{\psi}_\mu\left(\epsilon^{\mu\nu\rho\lambda}\gamma_5\gamma_\nu\partial_\rho+
 m\sigma^{\mu\lambda}\right)\psi_\lambda
\end{equation}

\noindent where $\epsilon^{\mu\nu\rho\lambda}$ is the Levi-Civita symbol, $\gamma^{5}=i\gamma^{0}\gamma^{1}\gamma^{2}\gamma^{3}$ and $\sigma^{\mu\lambda}=\frac{i}{2}\left[ \gamma^{\mu},\gamma^{\lambda} \right]$. The freedom represented by the parameter $A$ reflects invariance under rotations mixing the two spin-$1/2^{+}$ and $1/2^{-}$ sectors residing in the RS representation space besides spin-$3/2$ \cite{nath}-\cite{nap}.
By applying the Euler-Lagrange equations to (\ref{lrs}), the equation of motion are given by
\begin{equation}\label{em1}
 \left(\epsilon^{\mu\nu\rho\lambda}\gamma_{5}\gamma_{\nu}\partial_{\rho}+m\sigma^{\mu\lambda}\right)\psi_{\lambda}=0\,
\end{equation}
\noindent while (\ref{em1}) can be rewritten in a simpler form plus supplementary conditions
\begin{equation}\label{em3}
 (i\gamma^\mu\partial_\mu - m)\psi_\nu=0\,,
\end{equation}
\begin{equation}\label{v1}
 \gamma^\mu\psi_\mu=0\,,
\end{equation}
\begin{equation}\label{v2}
\partial^\mu\psi_\mu=0\,.
\end{equation}

\noindent These last results clearly show that the equations of motion (\ref{em1}) can be rewritten as a Dirac equation for the vector spinor plus supplementary conditions.
Therefore, the solution for this class of systems consists in searching for solutions for four Dirac-type equations.
It should not be forgotten, though, that the equations are not indeed independent because they have to satisfy the constraints (\ref{v1}) and (\ref{v2}).
These constraints are necessary to eliminate the redundant components of the vector spinor $\psi^{\mu}$.
These constraints, in general, do not hold in the presence of interactions.
Nevertheless, it is possible to derive the necessary number of subsidiary conditions for at least a selected class of interactions \cite{fierz}.
Furthermore, by multiplying equation (\ref{em3}) from the left by $\gamma^{\nu}$ and using equation (\ref{v1}), we obtain equation (\ref{v2}).
Therefore the \textquotedblleft gauge\textquotedblright condition (\ref{v2}) is not an independent
relation. The equations (\ref{em3}) and (\ref{v1}) are known as the RS equation \cite{greiner2}.

The RS equation can also be derived from the Lagrangian \cite{rs}
\begin{equation}\label{lrs2}
    \mathcal{L}^{RS}=\bar{\psi}_{\mu}\left( \slashed{\partial}+im \right)\psi^{\mu}-
    \frac{1}{3}\,\bar{\psi}_{\mu}\left( \partial^{\mu}\phi+\gamma^{\mu}\partial\cdot\psi \right)+
    \frac{1}{3}\,\bar{\psi}_{\mu}\gamma^{\mu}\left( \slashed{\partial}-im \right)\phi\,,
\end{equation}

\noindent where $\slashed{\partial}\equiv\gamma_{\mu}\partial^{\mu}$, $\phi\equiv\gamma_{\mu}\psi^{\mu}$ and $\partial\cdot\psi\equiv\partial_{\mu}\psi^{\mu}$.
In this case, the Euler-Lagrange equations of motion yield
\begin{equation}\label{em4}
    \left( \slashed{\partial}+im \right)\psi^{\mu}-\frac{1}{3}\,\left( \partial^{\mu}\phi+\gamma^{\mu}\partial\cdot\psi \right)+
    \frac{1}{3}\,\gamma^{\mu}\left( \slashed{\partial}-im \right)\phi=0\,.
\end{equation}

\noindent By contracting equation (\ref{em4}) with $\partial_{\mu}$ and $\gamma_{\mu}$ one obtains the supplementary conditions (\ref{v1}) and (\ref{v2}),
which allow equation (\ref{em4}) to reduce to the simpler form for the RS equation.
Due to simplicity once both equations are equivalent, we shall henceforth choose the equations (\ref{em3}) and (\ref{v1}) to describe a spin-$3/2$ particle.

%=================================
Now let us consider the system embedded in an electromagnetic field.
This is implemented via minimal coupling prescription, which consists in the replacement
\begin{equation}\label{am}
 \partial_\mu \rightarrow D_\mu=\partial_\mu +i q A_\mu\,,
\end{equation}
\noindent where $A_{\mu}$ is the $\mu$-th component of the four-vector
potential that represents
an external electromagnetic field and the charge $q=\epsilon_q|q|$,
where $\epsilon_q=+1(-1)$ corresponds to a particle with positive (negative) charge, respectively.
So, the RS equation with an electromagnetic field becomes
\begin{equation}\label{rsb}
  (i\gamma^\mu D_\mu - m)\psi_\nu=0 \quad \mathrm{with} \quad
 \gamma^\mu\psi_\mu=0\,.
\end{equation}

\noindent According to the previous discussion on the change of the subsidiary condition,
we can calculate the new condition by multiplying the first equation in (\ref{rsb}) from the left by $\gamma^{\nu}$ and using the second equation in (\ref{rsb}), leading to
\begin{equation}\label{v3}
   D^{\mu}\psi_{\mu}=0\, ,
\end{equation}

\noindent which is in agreement with the minimal coupling prescription given in equation (\ref{v2}).
This last result clearly shows that the minimal coupling prescription is mathematically consistent,
in contrast to the introduction of a Lorentz scalar potential,
which gets to a contradiction if we do the same in the RS equation
written in the form (\ref{em3}),
as addressed in \cite{velo,hortacsu,jasiak,blunden}.
To achieve the solutions of the RS equation embedded in an
electromagnetic field, we follow the
approach used in obtaining  the solutions of the Dirac equation in
\cite{melrose}. A similar approach was also used in \cite{greiner}.

Because of the reasons mentioned in the Introduction, we focus our
attention on considering the
external vector field to be a constant magnetic field.
Assuming the gauge $A_{\mu}=\delta_{\mu2}xB$, \textit{i.e.} $A_{\mu}=\left( 0,0,xB,0 \right)$
that yields a constant magnetic field transverse to the $xy$-plane, we have
 $\vec{\nabla}\cdot\vec{A}=0$ and
 $\vec{B}=\vec{\nabla}\times\vec{A}=B\hat{e}_{3}$. In this manner,
the covariant derivative becomes
\begin{equation}\label{dc}
  D_\mu=\partial_\mu -i\epsilon_q|q|Bx\delta_{\mu2}.
\end{equation}

\noindent As we have chosen the constant magnetic field to be transverse to the $xy$-plane,
we can apply the method of the separation of variables in the vector
spinor $\psi_{\mu}$, which can be
rewritten in the form
\begin{equation}
 \psi_{\mu}^{(\epsilon)}(\vec{x},t)=\phi_{\mu}^{(\epsilon)}(\vec{x})\mathrm{e}^{-i\epsilon Et}
=f_{\mu}^{(\epsilon)}(x)\mathrm{e}^{-i\epsilon Et+i \epsilon
  p_{y}y+i\epsilon p_{z}z}\, ,
\end{equation}
\noindent where we define $E$ as positive, and $\epsilon=+1(-1)$ corresponds to the states of positive (negative) energy, respectively.

Taking advantage of the solutions for the Dirac equation, we can consider the solution for $\epsilon_q=+1$ as
\begin{equation}\label{fqp1}
   f^{(\epsilon)}_{\mu}=
   \left(
     \begin{array}{c}
       f_{\mu1}(x) \\
       f_{\mu2}(x) \\
       f_{\mu3}(x) \\
       f_{\mu4}(x) \\
     \end{array}
   \right)=
   \left(
   \begin{array}{c}
     C^{(\epsilon)}_{\mu1}v_{n}(\xi) \\
     C^{(\epsilon)}_{\mu2}v_{n-1}(\xi) \\
     C^{(\epsilon)}_{\mu3}v_{n}(\xi) \\
     C^{(\epsilon)}_{\mu4}v_{n-1}(\xi)
   \end{array}
   \right) ,
\end{equation}

\noindent and for $\epsilon_q=-1$ as
\begin{equation}\label{fqm1}
   f^{(\epsilon)}_{\mu}=
   \left(
     \begin{array}{c}
       f_{\mu1}(x) \\
       f_{\mu2}(x) \\
       f_{\mu3}(x) \\
       f_{\mu4}(x) \\
     \end{array}
   \right)=
   \left(
   \begin{array}{c}
     C^{(\epsilon)}_{\mu1}v_{n-1}(\xi) \\
     C^{(\epsilon)}_{\mu2}v_{n}(\xi) \\
     C^{(\epsilon)}_{\mu3}v_{n-1}(\xi) \\
     C^{(\epsilon)}_{\mu4}v_{n}(\xi)
   \end{array}
   \right) ,
\end{equation}

\noindent where
\begin{equation}\label{vn}
    v_{n}(\xi)=\frac{1}{\sqrt{\pi^{1/2}2^{n}n!}}\,H_{n}(\xi)\mathrm{e}^{-\xi^2/2} ,
\end{equation}

\noindent is the solution for the one-dimensional harmonic oscillator, where energy takes the form
\begin{equation}\label{energy}
    E=\sqrt{p_{z}^{2}+m^{2}+2n|q|B}
\end{equation}

\noindent and the coefficients $C_{\mu}^{(\epsilon)}$ satisfy the relation
 \begin{equation}
 \left (
 \begin{array}{c}
  C^{(\epsilon)}_{\mu 1} \\
  C^{(\epsilon)}_{\mu 2} \\
  C^{(\epsilon)}_{\mu 3} \\
  C^{(\epsilon)}_{\mu 4}
 \end{array}
 \right )
 =
 \left (
 \begin{array}{c}
  1 \\
  0 \\
  \frac{\epsilon p_z}{\epsilon E+m} \\
  -\frac{i\epsilon_q p_n}{\epsilon E+m}
 \end{array}
 \right )C^{(\epsilon)}_{\mu 1}
 +
 \left (
 \begin{array}{c}
  0 \\
  1 \\
  \frac{i\epsilon_q p_n}{\epsilon E+m} \\
  -\frac{\epsilon p_z}{\epsilon E+m}
 \end{array}
 \right )C^{(\epsilon)}_{\mu 2}.
\end{equation}

\noindent We have now to consider the subsidiary conditions. We can define the operators
\begin{equation}\label{op1}
    \hat{O}_1=i ( \epsilon p_y -\epsilon_q|q|Bx +
    \frac{\partial}{\partial x} ) = i(|q|B)^{1/2}(-\epsilon_q \xi +
    \frac{\partial}{\partial \xi})\, ,
\end{equation}
\begin{equation}\label{op2}
    \hat{O}_2=i (-\epsilon p_y +\epsilon_q|q|Bx +
    \frac{\partial}{\partial x} ) = i(|q|B)^{1/2}( \epsilon_q \xi +
    \frac{\partial}{\partial \xi}) ,
\end{equation}

\noindent noticing that
\begin{equation}\label{op1v}
    \hat{O}_1v_n=
\left\{
\begin{array}{c}
-ip_{n+1}v_{n+1}\quad  \mathrm{if}\quad \epsilon_q= 1 , \\
ip_{n}v_{n-1}  \qquad  \mathrm{if}\quad \epsilon_q=-1 ,
\end{array}
\right.
\end{equation}
\noindent and
\begin{equation}\label{op2v}
    \hat{O}_2v_n=
\left\{
\begin{array}{c}
ip_{n}v_{n-1}\qquad  \mathrm{if}\quad \epsilon_q= 1 ,\\
-ip_{n+1}v_{n+1}  \quad  \mathrm{if}\quad \epsilon_q=-1 ,
\end{array}
\right.
\end{equation}

\noindent where $p_{n}=\sqrt{2n|q|B}$. Hence, the condition (\ref{v3})(\textit{i.e.}  $D^\mu\psi_\mu=0$) becomes
\begin{equation}\label{vc1}
\begin{split}
&
-\epsilon E
  \left (
 \begin{matrix}
  C^{(\epsilon)}_{01} v_{n_q}(\xi) \\
  C^{(\epsilon)}_{02} v_{m_q}(\xi) \\
  C^{(\epsilon)}_{03} v_{n_q}(\xi) \\
  C^{(\epsilon)}_{04} v_{m_q}(\xi)
 \end{matrix}
 \right )
+\frac{1}{2}
 \left(
 \begin{matrix}
  p_{n_q+1} (i\epsilon_qC^{(\epsilon)}_{x1}-C^{(\epsilon)}_{y1}) v_{n_q+1}(\xi) \\
  p_{m_q+1} (i\epsilon_qC^{(\epsilon)}_{x2}-C^{(\epsilon)}_{y2}) v_{m_q+1}(\xi) \\
  p_{n_q+1} (i\epsilon_qC^{(\epsilon)}_{x3}-C^{(\epsilon)}_{y3}) v_{n_q+1}(\xi) \\
  p_{n_q+1} (i\epsilon_qC^{(\epsilon)}_{x4}-C^{(\epsilon)}_{y4}) v_{m_q+1}(\xi)
 \end{matrix}
 \right)
\\
&
+\frac{1}{2}
 \left(
 \begin{matrix}
  p_{n_q} (-i\epsilon_qC^{(\epsilon)}_{x1}-C^{(\epsilon)}_{y1}) v_{n_q-1}(\xi) \\
  p_{m_q} (-i\epsilon_qC^{(\epsilon)}_{x2}-C^{(\epsilon)}_{y2}) v_{m_q-1}(\xi) \\
  p_{n_q} (-i\epsilon_qC^{(\epsilon)}_{x3}-C^{(\epsilon)}_{y3}) v_{n_q-1}(\xi) \\
  p_{m_q} (-i\epsilon_qC^{(\epsilon)}_{x4}-C^{(\epsilon)}_{y4}) v_{m_q-1}(\xi)
 \end{matrix}
 \right)
+\epsilon p_z
 \left(
 \begin{matrix}
  C^{(\epsilon)}_{z1} v_{n_q}(\xi) \\
  C^{(\epsilon)}_{z2} v_{m_q}(\xi) \\
  C^{(\epsilon)}_{z3} v_{n_q}(\xi) \\
  C^{(\epsilon)}_{z4} v_{m_q}(\xi)
 \end{matrix}
 \right)=0,
\end{split}
\end{equation}

\noindent where we define $n_{q}=n$ ($n_{q}=n-1$) and $m_{q}=n-1$ ($m_{q}=n$) when the charge is positive (negative), respectively.

Finally, from the restriction (\ref{v1})(\textit{i.e.} $\gamma^\mu\psi_\mu=0$), we obtain
\begin{equation}\label{vc2}
    \left(
    \begin{array}{c}
      C^{(\epsilon)}_{01}v_{n_q}(\xi) \\
      C^{(\epsilon)}_{02}v_{m_q}(\xi) \\
      -C^{(\epsilon)}_{03}v_{n_q}(\xi) \\
      -C^{(\epsilon)}_{04}v_{m_q}(\xi)
    \end{array}
    \right)+
    \left(
    \begin{array}{c}
      C^{(\epsilon)}_{x4}v_{m_q}(\xi) \\
      C^{(\epsilon)}_{x3}v_{n_q}(\xi) \\
      -C^{(\epsilon)}_{x2}v_{m_q}(\xi) \\
      -C^{(\epsilon)}_{x1}v_{n_q}(\xi)
    \end{array}
    \right)+i
    \left(
    \begin{array}{c}
      -C^{(\epsilon)}_{y4}v_{m_q}(\xi) \\
      C^{(\epsilon)}_{y3}v_{n_q}(\xi) \\
      C^{(\epsilon)}_{y2}v_{m_q}(\xi) \\
      -C^{(\epsilon)}_{y1}v_{n_q}(\xi)
    \end{array}
    \right)+
    \left(
    \begin{array}{c}
      C^{(\epsilon)}_{z3}v_{n_q}(\xi) \\
      -C^{(\epsilon)}_{z4}v_{m_q}(\xi) \\
      -C^{(\epsilon)}_{z1}v_{n_q}(\xi) \\
      C^{(\epsilon)}_{z2}v_{m_q}(\xi)
    \end{array}
    \right)=0\,.
\end{equation}

\noindent From the system of equations (\ref{vc1}) and (\ref{vc2}), we conclude that for $n\geq2$ only four coefficients are independent
meaning that the states of positive or negative energy are quadruply degenerate. On the other hand, for $n=1$ (considering $\epsilon_{q}=-1$) from the system of equations (\ref{vc1}) and (\ref{vc2}) ($v_{-1}=0$), we obtain that only three coefficients are independent meaning that the energy state for $n=1$ is triply degenerate. Finally, for $n=0$ (considering $\epsilon_{q}=-1$) from equation (\ref{fqm1}) we obtain that $C^{(\epsilon)}_{\mu1}=C^{(\epsilon)}_{\mu3}=0$, and in this case from (\ref{vc1}) and (\ref{vc2}) ($v_{-2}=v_{-1}=0$) we conclude that only two coefficients are independent, meaning that the energy state for $n=0$ is doubly degenerate. This degeneracy of the energy state can be associated with a spin label, as well as
\begin{equation}\label{sl}
    n=l-\frac{s}{2}\,\epsilon_{q}+\frac{1}{2} \,,
\end{equation}

\noindent where $l=0,1,2,\ldots$ and $s=\pm1,\pm3$. Therefore, the
degeneracy factor for spin $3/2$ particles are given by
\begin{equation}\label{dge}
    g_{n}=4-\delta_{n1}-2 \delta_{n0}\,.
\end{equation}

\noindent Explicitly, we obtain $g_{0}=2$ ($n=0$) for $l=0$, $s=-1$
and $l=1$, $s=-3$. We obtain $g_{1}=3$ ($n=1$) for $l=0$, $s=+1$;
$l+1$, $s=-1$ and $l=2$, $s=-3$. And finally, we obtain $g_{n}=2s+1=4$
($n\geq2$) for $s=3/2$. At this stage, it is useful to mention that
independent coefficients may be completely calculated following an
analogue procedure to the one described in the section 5.2 of the
Ref. \cite{melrose2}, i.e, using an appropriate helicity operator for
spin $3/2$ particles \cite{sarkar}, but these calculations are beyond the scope of this work.

\section{Final discussions}

In the present paper, we revisit the Rarita-Schwinger equation and
include the electromagnetic interaction through the minimum coupling
scheme. We assume a gauge where the magnetic field transverse to the
$xy$-plane is constant. This choice leads to Dirac-type solutions,
which can consistently avoid noncausality problems.
 We must observe that the energy given in eq.(\ref{energy}) may assume
complex values for $n\geq 1$ if $B>m^2/2n|q|$,
but this is a very strong magnetic field ($|eB|\sim10^{20}$G), much stronger than the estimates
for the typical values of the fields produced in heavy ion collisions \cite{rhic} for example. 
Moreover,
we can clearly see that the occupation of the Landau levels by
  spin $3/2$ particles is quite different from the occupation of
  spin $1/2$ particles.
While systems composed of spin $1/2$ particles subject to
magnetic fields can accommodate only one particle in the lowest level
and two particles in the other levels, systems with spin $3/2$ particles
can accommodate $2$ particles in the first level, $3$ particles in the
second one and $4$ particles in the remaining levels. This fact has a
great  influence in studies involving equations of state,
density of particles, scattering amplitudes and other
calculations of interest for  nuclear astrophysics and physics
  of heavy ion collisions mentioned in the Introduction.

\section*{ACKNOWLEDGMENTS}

This work was partially supported by CNPq, CAPES and FAPESC (Brazil).
We thank Prof. Melrose for carefully reading this manuscript.

\end{document}